\documentclass[12pt]{article}
\usepackage{graphics,amssymb,latexsym,amsfonts,graphics}

\usepackage{amsmath, amssymb, graphics}

\begin{document}
\thispagestyle{empty}

\begin{center}
{\textbf{\Large{ Effect of axial electric field on the binding energy of a
 shallow hydrogenic impurity in a Quantum Well Wire }}}\\
I F I  $Mikhail^{1}$ and T G $Emam^{1,2}$\\
$^1$ Department of Mathematics, Faculty of science, Ain Shams
university, Abbassia, Cairo, Egypt.\\
$^2$ Department of Mathematics, the German university in
Cairo-GUC, New Cairo city,  Main Entrance Al Tagamoa Al Khames,
Egypt.
\end{center}
Abstract:  We present a study of the effect of an electric field
on the binding energy of a shallow hydrogenic impurity in
$GaAs/Ga_{1-x}Al_{x}As$ quantum well wires. The wire is considered
to be of length $L$ and radius $R$ and the electric field $F$ is
applied
along the $z-$axis, the axis of the wire.\\
\section{Introduction}
   The development of the epitaxial crystal growth
  techniques such as molecular beam epitaxy and metal-organic
  chemical vapor deposition has made the growth of the
  quasi-two-dimensional (quantum well) , quasi-one- dimensional
  (quantum well wire), or quasi-zero-dimensional (quantum dot) become possible [1,2,3].
  \vskip 0.5 cm Several studies have investigated the optical and transport
  properties based on the calculated electronic structure of these
  systems in the presence of shallow impurities (see for example references [4-9]).
  \vskip 0.2 cm Regarding the quantum well wire(QWW), the binding
  energy of hydrogenic impurities in a QWW has been calculated as
  a function of the wire radius in case of no fields applied[7,8]
. Also, Branis et al [10] considered the case of a magnetic field
  applied parallel to the wire axis. They found that for a given
  value of the magnetic field, the binding energy is
  larger than that of  the zero-field case.
  \vskip 0.2 cm In this work we have considered  the case of an electric
  field applied along  the wire axis, we have calculated the
  binding energy of the ground state of a hydrogenic impurity
  located at the wire axis. A variational approach has been utilized with a  trail wave function
   containing  a hydrogenic part. The other part of the variational function which represents the wave function
   in the absence of the impurity has been calculated analytically in terms of Bessel and Airy functions.
\section{Electron Eigenstates}
 Consider a quantum well  wire of length L and radius R (of dimension comparable to de Broglie's
 wavelength), with an electric field of strength  F  applied along the axis of the
 wire .
 The Hamiltonian for such a system is given by:
\begin{equation}
H = \frac{-\hbar^2}{2m_e} \nabla^2 + |e| Fz + V(\rho, z),
\end{equation}
where
\begin{equation}
V(\rho, z) = \begin{cases}0, & \rho \leq R \ and \ |z| \leq
\frac{L}{2} \cr
 & \cr
\infty, & \rho > R \ or \ |z| > \frac{L}{2}, \end{cases}.
\end{equation}
 We have  used  the cylindrical
coordinates such that the origin is taken at the mid point of the
wire  axis, which is considered as z-axis \vskip 0.2cm The
Schr\"{o}dinger equation takes the form
\begin{equation}
\frac{1}{\rho}\frac{\partial}{\partial
\rho}\bigl(\rho\frac{\partial \psi}{\partial
\rho}\bigr)+\frac{1}{\rho^2}\frac{\partial^2 \psi}{\partial
\phi^2}+ \frac{\partial^2 \psi}{\partial z^2}-\frac{2m_e|e|
F}{\hbar^2} z \psi= \frac{-2m_eE_0}{\hbar^2}\psi .
\end{equation}
Assuming that
\begin{equation}
\psi = \Phi (\phi) \zeta (z) {\Re}(\rho ),
\end{equation}
and noting that the problem is rotational invariant $(\Phi
(\phi)=e^{im\phi})$, then Eq. (3) takes the form
\begin{equation}
\frac{1}{\Re \rho}\frac{d}{d \rho}\bigl(\rho\frac{d \Re}{d
\rho}\bigr)-\frac{m^2}{\rho^2}+\frac{1}{\zeta}\frac{d^2
\zeta}{dz^2}-\frac{2m_e|e| F}{\hbar^2} z= \frac{-2m_eE}{\hbar^2}.
\end{equation}
The radial part is
\begin{equation}
\Re(\rho) = J_m(k_{nm}\rho),
\end{equation}
where $J_m$ is the ordinary Bessel function . \vskip 0.2cm
Accordingly the axial part $\zeta(z)$ is found to satisfy the
equation
\begin{equation}
\frac{1}{\zeta}\frac{d^2 \zeta}{dz^2}-\frac{2m_e|e| F}{\hbar^2} z
+ \frac{2m_eE_0}{\hbar^2}- k_{nm}^2 = 0.
\end{equation}
To solve this equation, we assume that
\begin{equation}
\tilde{z} = G^{\frac{-2}{3}} \Bigl(Gz - \frac{2m_eE_0}{\hbar^2}
+k_{nm}^2\Bigr),
\end{equation}
where
\begin{equation}
G = \frac{2m_e|e| F}{\hbar^2}.
\end{equation}
Substituting from Eq. (8) into Eq. (7) we obtain the equation
\begin{equation}
\frac{d^2 \zeta}{d\tilde{z}^2}= \tilde{z}\zeta ,
\end{equation}
which is Airy's equation  The solution of Eq. $(10)$ is a linear
combination of  the Airy functions $Ai(\tilde{z})$ and
$Bi(\tilde{z})$. (Cetina and Montenegro [12])Thus, the final form
of $\psi(\underline{r})$ is taken as
\begin{equation}
\psi_{nm}(\underline{r})=Ne^{im\phi}J_m(k_{nm}\rho)\Bigl[Bi(\tilde{z})
- \frac{Bi(\xi_l)}{Ai(\xi_l)}Ai(\tilde{z})\Bigr],
\end{equation}
where $\xi_l = \tilde{z}$ $(z= \frac{-L}{2})$. The form of
$\zeta(z)$ in Eq. (11) has been chosen so that $\psi$ vanishes at
$z=\frac{-L}{2}$. We still need to satisfy the boundary conditions
(Vanishing of the wave function at the boundaries)
\begin{equation}
\psi (\rho =R)=0, \hspace{1.0cm} \psi(z=\frac{L}{2}) = 0.
\end{equation}
The first gives
\begin{equation}
J_m(k_{nm}R) = 0.
\end{equation}
For the ground state ($m=0$, $n=1$), which is  considered in this
work, the above equation implies
\begin{equation}
k_{10}R = 2.4048,
\end{equation}
where the R.H.S. is the first root of $J_0(x)=0$. Also, the second
condition in Eq.(12) yields
\begin{equation}
Bi(\xi_h)Ai(\xi_l) = Bi(\xi_l)Ai(\xi_h),
\end{equation}
where $\xi_h = \tilde{z}$ $(z= \frac{L}{2})$. \vskip 0.2cm The
energy states can consequently be obtained by solving the
transcendental Eq. (15). The energy of the ground state will be
denoted by $E_0$.
\subsection{Impurity Eigenstates}
We consider the problem of a hydrogenic impurity located at $\rho
= z=0$. The trial wave function for the ground state is taken as
\begin{equation}
\psi = N J_0(k_{10}R)(Bi(\tilde{z})-\frac{Bi(\xi_l)}{Ai(\xi_l)}
Ai(\tilde{z}))e^{-\lambda\sqrt{\rho^2+z^2}}.
\end{equation}
The normalization condition  leads to
\begin{equation}
N^{-2} =2\pi \int_{\rho
=0}^R\int_{z=\frac{-L}{2}}^{\frac{L}{2}}d\rho dz \
 \rho J_0^2(k_{10}R)\Bigl(Bi(\tilde{z})-\frac{Bi(\xi_l)}{Ai(\xi_l)}Ai(\tilde{z})\Bigr)^2
 e^{-2\lambda\sqrt{\rho^2+z^2}}
\end{equation}
The potential energy $\langle V \rangle$ is
\begin{eqnarray}
\nonumber \langle V \rangle &=& \langle \psi
|\frac{-e^2}{\epsilon_0 \sqrt{\rho^2+z^2}} |
\psi\rangle\\
\nonumber &=& 2\pi N^2
\bigl(\frac{-e^2}{\epsilon_0}\bigr)\int_{\rho =0}^R
\int_{z=\frac{-L}{2}}^{\frac{L}{2}}d\rho dz \frac{\rho J_0^2(k_{10}\rho)}{\sqrt{\rho^2+z^2}}\\
& \times &
\Bigl(Bi(\tilde{z})-\frac{Bi(\xi_l)}{Ai(\xi_l)}Ai(\tilde{z})\Bigr)^2
e^{-2\lambda\sqrt{\rho^2+z^2}}.
\end{eqnarray}
Putting the potential energy in unit of Ryrderg $R_b$ and length
in unit of Bohr radius $a_B$, $\langle V \rangle$ takes the form
\begin{eqnarray}
\nonumber
\langle V \rangle &=& -4\pi N^2 \int_{\rho =0}^R \int_{z=\frac{-L}{2}}^{\frac{L}{2}}
d\rho dz \frac{\rho e^{-2\lambda\sqrt{\rho^2+z^2}}}{\sqrt{\rho^2+z^2}}\\
&\times&
J_0^2(k_{10}\rho)\Bigl(Bi(\tilde{z})-\frac{Bi(\xi_l)}{Ai(\xi_l)}Ai(\tilde{z})\Bigr)^2.
\end{eqnarray}
To calculate the kinetic energy $\langle T \rangle$ we first
calculate $\nabla^2 \psi$ as
\[ \psi = \psi_1 \ \psi_2, \]
where
\[\psi_1 = e^{-\lambda\sqrt{\rho^2+z^2}} \]
and \[ \psi_2
=J_0(k_{10}\rho)\Bigl(Bi(\tilde{z})-\frac{Bi(\xi_l)}{Ai(\xi_l)}Ai(\tilde{z})\Bigr).
\] Hence
\begin{equation}
\nabla^2 \psi = (\nabla^2 \psi_1)\psi_2 + \psi_1 (\nabla^2 \psi_2)
+ 2(\nabla \psi_1)(\nabla \psi_2),
\end{equation}
but
\begin{equation}
(\nabla^2 -G z)\psi_2=\frac{-2m_eE_0}{\hbar^2} \psi_2
\end{equation}
and
\begin{equation}
\nabla^2 \psi_1=\Bigl(\lambda^2 -
\frac{2\lambda}{\sqrt{\rho^2+z^2}}\Bigr)\psi_1.
\end{equation}
We then  evaluate the term
\begin{eqnarray}
\nonumber
2 \langle \psi \mid  \nabla \psi_1 \nabla \psi_2 \rangle &=& 2 \int d\phi dz d \rho \rho (\nabla \psi_1)(\nabla \psi_2)\psi_1^{\ast} \psi_2^{\ast}\\
\nonumber
&=& \frac{1}{2} \int d\phi dz d \rho \ \rho \Bigl(\frac{\partial \psi_1^2}{\partial \rho}\frac{\partial \psi_2^2}{\partial \rho}+\frac{\partial \psi_1^2}{\partial z}\frac{\partial \psi_2^2}{\partial z}\Bigr)\\
\nonumber
&=& \int d\phi dz d \rho \ \rho \Bigl(\frac{-\lambda \rho}{\sqrt{\rho^2 +z^2}} \psi_1^2\Bigr) \frac{\partial \psi_2^2}{\partial \rho}\\
&+& \int d\phi dz d \rho \ \rho \Bigl(\frac{-\lambda
z}{\sqrt{\rho^2 +z^2}} \psi_1^2\Bigr) \frac{\partial
\psi_2^2}{\partial z}
\end{eqnarray}
Integrating the first term by parts with respect to $\rho$ and the
second term by parts with respect to $z$ and using the boundary
conditions (vanishing of $\psi_2$ at the boundary) we get
\begin{equation}
2 \langle \psi \mid  \nabla \psi_1 \nabla \psi_2 \rangle =
-2\lambda^2 \int d\phi dz d \rho  \ \rho \psi_1^2 \psi_2^2 +
2\lambda \int d\phi dz d \rho \frac{\rho}{\sqrt{\rho^2 +z^2}}
\psi_1^2\psi_2^2.
\end{equation}
Using Eq. (21),Eq.(22), and Eq. (24), we get
\begin{equation}
  \frac{-\hbar^2}{2m_e} \langle \psi| \nabla^2 -G z| \psi \rangle = E_0 + \frac{\hbar^2 \lambda^2}{2m_e}
\end{equation}

Putting Eq. $(25) $ in units of Rydberg $R_b$, and length in units
of Bohr radius $a_{B}$ and adding to $\langle V \rangle $, the
total energy is
\begin{equation}
\langle H(R, L, F) \rangle = E_0 + \lambda^2 +
\frac{4A}{\frac{dA}{d\lambda}}
\end{equation}
where
\begin{equation}
A = \int_{\rho =0}^R \int_{z=\frac{-L}{2}}^{\frac{L}{2}} d\rho dz
\frac{\rho e^{-2\lambda\sqrt{\rho^2+z^2}}}{\sqrt{\rho^2+z^2}}
 J_0^2(K_{10}\rho)\Bigl(Bi(\tilde{z})-\frac{Bi(\xi_l)}{Ai(\xi_l)}Ai(\tilde{z})\Bigr)^2.
\end{equation}
 Minimizing $\langle H(R, F, L) \rangle$ with respect to $\lambda$,
we obtain the total energy, and the binding energy is thus
\begin{equation}
 E_b = E_0 - \langle H(R, F, L) \rangle =-\lambda^2-\frac{4
A}{\frac{d A}{d \lambda}}
\end{equation}

\section{ Results and Discussions} \vskip 0.2cm
The variation of the binding energy with the electric field
strength F and the wire radius  R are given in figures $(1)$ and
$(2)$ respectively. The results displayed in Figure$(1)$ show that
for  a wire of fixed dimensions L and R the binding energy
decreases as the strength of the applied electric field increases.
This is due to the fact that the increase of the electric field
causes the electron to be less confined to the impurity and
accordingly reduces the binding energy. Also the results in
figures $(1) , (2)$ indicate that for a fixed electric field  the
binding energy increases as the wire radius decreases until it
diverges as $R\rightarrow 0$. The divergence of the binding energy
occurs since the infinite confining potential of the well forces
the electron and the impurity to become very close as
$R\rightarrow 0$. The same behavior has been previously pointed
out in the absence of the electric field (ref[7]).
\section{Conclusions}
 The binding energy of the hydrogenic donor impurity decreases as
 the strength of the applied electric field increases. It behaves
 with the variation of the wire radius in a similar manner to the
 case of no electric field applied.

\end{document}